\documentclass[conference]{IEEEtran}
\usepackage[utf8]{inputenc}
\usepackage{float}

\usepackage[detect-all]{siunitx}
\DeclareSIUnit{\belmilliwatt}{Bm}
\DeclareSIUnit{\dBm}{\deci\belmilliwatt}
\DeclareSIUnit{\belisotropic}{Bi}
\DeclareSIUnit{\dBm}{\deci\belisotropic}
\DeclareSIUnit{\bit}{bit}
\usepackage{xcolor}
\usepackage{amsmath}
\usepackage{array,tabularx,calc}
\usepackage{mathtools}
\usepackage{amsfonts}
\usepackage{graphicx}
\usepackage{amssymb}
\usepackage[caption=false]{subfig}
\usepackage{etoolbox}

\ifCLASSINFOpdf
\else
\fi

\begin{document}
%

\title{A Routing Metric for Inter-flow Interference-aware Flying Multi-hop Networks}

\author{\IEEEauthorblockN{André Coelho, Eduardo Nuno Almeida, José Ruela, Rui Campos, Manuel Ricardo}
	\IEEEauthorblockA{INESC TEC and Faculdade de Engenharia, Universidade do Porto, Portugal\\
		\{andre.f.coelho, eduardo.n.almeida, jose.ruela, rui.l.campos, manuel.ricardo\}@inesctec.pt}}


\maketitle

\begin{abstract}
	The growing demand for broadband communications anytime, anywhere has paved the way to the usage of Unmanned Aerial Vehicles (UAVs) for providing Internet access in areas without network infrastructure and enhancing the performance of existing networks. However, the usage of Flying Multi-hop Networks (FMNs) in such scenarios brings up significant challenges concerning network routing, in order to permanently provide the Quality of Service expected by the users. The problem is exacerbated in crowded events, where the FMN may be formed by many UAVs to address the traffic demand, causing inter-flow interference within the FMN. 
	Typically, estimating inter-flow interference is not straightforward and requires the exchange of probe packets, thus increasing network overhead.
		
	The main contribution of this paper is an inter-flow interference-aware routing metric, named I2R, designed for centralized  routing in FMNs with controllable topology. I2R does not require any control packets and enables the configuration of paths with minimal Euclidean distance formed by UAVs with the lowest number of neighbors in carrier-sense range, thus minimizing inter-flow interference in the FMN. 
	Simulation results show the I2R superior performance, with significant gains in terms of throughput and end-to-end delay, when compared with state of the art routing metrics.
		
\end{abstract}

\IEEEpeerreviewmaketitle

\section{Introduction}
\begin{figure}
	\setlength\abovecaptionskip{-0.1\baselineskip}
	\centering
	\includegraphics[scale=0.18]{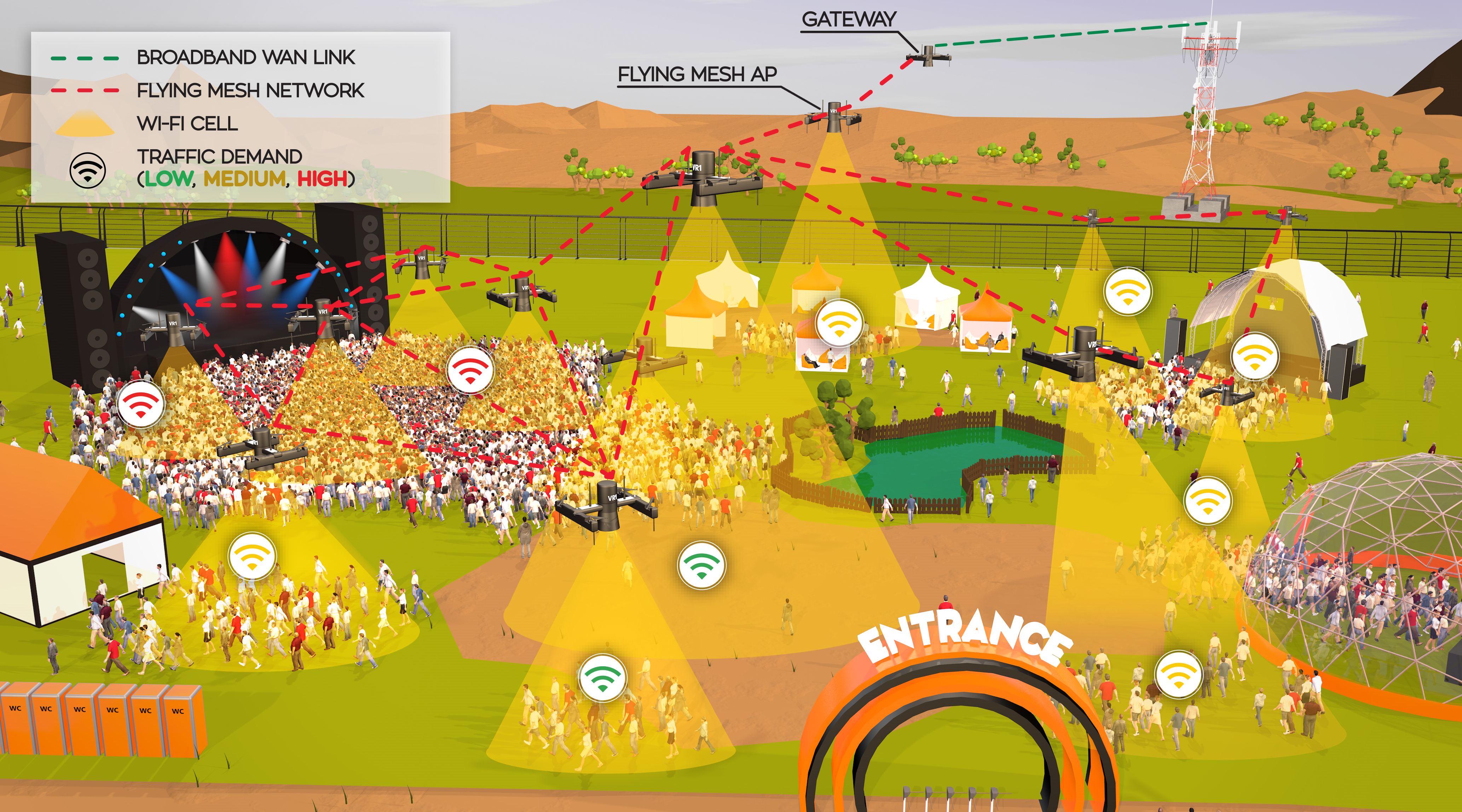}
	\caption{A Flying Multi-hop Network deployed in a music festival to provide Internet connectivity to the terminals on the ground.}
	\label{fig:wise-concept}
\end{figure}

In the last years, Unmanned Aerial Vehicles (UAVs) have been used in several applications, including environmental monitoring, border surveillance, emergency assistance, search, rescue, and payload transport~\cite{Hayat2016}. Meanwhile, the growing demand for broadband communications anytime, anywhere has paved the way to the usage of UAVs to 1) provide Internet access in areas without network infrastructure and 2) enhance the performance of existing networks~\cite{Almeida2018}. The ability to operate anywhere, their mobility and hovering capabilities, and their growing payload make UAVs viable platforms to carry network hardware, including Wi-Fi Access Points and Long-Term Evolution (LTE) Base Stations. In this sense, a swarm of UAVs can be deployed to form a mobile and physically reconfigurable aerial network infrastructure covering a large area, where the UAVs cooperatively forward the traffic to the Internet along multi-hop paths, as proposed by the WISE project~\cite{Almeida2018}, whose concept is illustrated in Fig.~\ref{fig:wise-concept}. However, such scenarios bring up significant challenges concerning network routing, in order to permanently provide the Quality of Service (QoS) expected by the users, including always-on broadband Internet connectivity, even when the UAVs are moving and the Flying Multi-hop Network (FMN) topology is being reconfigured.\looseness=-1 

Setting up a multi-hop wireless network with broadband links has always been a challenging task. Due to the transmission through shared wireless channels, which characterizes IEEE 802.11-based multi-hop networks, the interference caused by the traffic being transmitted by neighboring nodes must be taken into account. This type of interference is commonly named inter-flow interference. The problem is exacerbated in crowded events, such as fun parks, parades, and festivals, where a large number of terminals need to be served, requiring an FMN with many UAVs close to each other to address the traffic demand. However, typically, estimating inter-flow interference is not straightforward, since it depends on the radio propagation environment and the nodes in carrier-sense range.\looseness=-1  

In~\cite{Coelho2018}, the authors propose RedeFINE, a centralized routing solution for FMNs that uses the Euclidean distance between the UAVs as routing metric to define in advance the forwarding tables and the instants they shall be updated in the UAVs. RedeFINE takes advantage of the holistic knowledge provided by a Central Station (CS), which is responsible for defining the locations of the UAVs to meet the traffic demand from the users on the ground. However, in its current version, RedeFINE is not suitable for highly dense networks, since the interference between neighboring UAVs is not considered. 

The main contribution of this paper is an inter-flow interference-aware routing  metric, named I2R, designed for centralized routing in FMNs with controllable topology. I2R does not require any control packets and enables the configuration of paths with minimal Euclidean distance formed by UAVs with the lowest number of neighbors in carrier-sense range, thus minimizing inter-flow interference in the FMN. I2R advances the state of the art by avoiding the usage of control packets for neighbor discovery and interference estimation. Due to the highly dynamic behaviour of FMNs, which induce frequent changes in the quality of the radio links and in the network topology, existing routing solutions need to constantly flood the network with route discovery packets, thus reducing the bandwidth available for data traffic. In I2R, since a strong Line-of-Sight (LoS) component characterizes the communications links between UAVs flying dozens of meters above the ground, we assume the wireless links can be modeled by the Friis path loss model, which is used to estimate the Signal-to-Noise Ratio (SNR) and identify the neighbors in carrier-sense range without any signaling.

The performance of I2R is evaluated by means of ns-3 simulation, considering a new version of RedeFINE that uses I2R as the routing metric. The results obtained for throughput and end-to-end delay allow to conclude the superior performance of I2R, when compared with state of the art routing metrics that use real measurements to estimate the link quality. 

The rest of the paper is organized as follows. Section~\ref{sec:soa} 
presents the state of the art on routing metrics for wireless networks. Section~\ref{sec:system-model} 
defines the system model, including the network model and the interference model used. Section~\ref{sec:concept} presents the concept of the I2R routing metric. Section~\ref{sec:performance-evaluation} details the performance evaluation, including the simulation setup, the simulation scenarios, the performance metrics studied, and the simulation results.
Finally, Section~\ref{sec:conclusions} presents the conclusions and directions for future work.

\section{State of the Art\label{sec:soa}}
The ability to provide the QoS expected by the users is an important requirement in FMNs that are used to provide Internet access in crowded events. Existing routing solutions for FMNs are based on the protocols employed in Mobile Ad Hoc Networks (MANETs) and Vehicular Ad Hoc Networks (VANETs). In such networks, several works have been performed to meet the QoS requirements, and many of them have proposed different routing metrics. In the following, the most relevant ones are presented. 

Expected Transmission Count (ETX)~\cite{Couto2005} aims at estimating the number of transmissions required to successfully deliver a packet through a wireless link, by capturing the Packet Loss Ratio and link length. Similarly, Expected Transmission Time (ETT)~\cite{draves2004} aims at estimating the MAC layer duration to successfully transmit a packet through a wireless link. ETT is an improvement of ETX since it takes into account the quality and capacity of the wireless links. However, neither ETX nor ETT include the effect of interference over the wireless links. Weighted Cumulative Expected Transmission Time (WCETT)~\cite{draves2004} was the first multi-channel routing metric that included the effect of interference. WCETT improves ETT by considering the intra-flow interference, aiming at selecting paths composed of links operating on as many different channels as possible. However, WCETT is not isotonic, which means that the order of the weights of two paths is not preserved if they are appended or prefixed by a third path; hence, WCETT does not allow to ensure paths without loops. Additionally, WCETT does not capture inter-flow interference, which makes it not appropriate for highly dense networks. Metric of Interference and Channel-switching (MIC)~\cite{Yang2005} represents a step forward to WCETT, since it is isotonic and takes into account intra-flow and inter-flow interference; nevertheless, MIC is designed for static networks, since it does not consider that interference can vary along the time, due to changes in the signal strength, and the amount of traffic transmitted by the nodes. The Interference AWARE (iAWARE)~\cite{Subramanian2006} metric uses SNR and Signal-to-Interference plus Noise Ratio (SINR) to estimate inter-flow and intra-flow interference. However, iAWARE is not isotonic. The previous metrics are adopted by most of the routing protocols used in wireless networks. An exception is the Better Approach To Mobile Ad-hoc Networking (B.A.T.M.A.N.)~\cite{neumann2008} routing protocol, which introduced its own routing metric. In B.A.T.M.A.N., nodes broadcast originator messages (OGMs) to inform the neighboring nodes about their existence. OGMs allow to infer the quality of the wireless links and the congestion of the network. The next-hop of each node to reach the source of the received OGMs is the one that delivered the highest number of OGMs during a period of time.

All the previous metrics use probe packets to get the measurements they need, which may cause high overhead and may not be scalable for large networks. Additionally, in FMNs, inter-flow interference is not a local concept, being related to all the interfering nodes along a path. Then, a routing metric fed by a holistic view of the network is desirable to be considered. 

\section{System Model\label{sec:system-model}}
The system model is presented in this section, including the network model and the interference model used.

\subsection{Network Model}
The FMN consists of $N$ UAVs that are controlled by a CS. The CS is in charge of 1) defining the FMN topology, so that the UAVs meet the traffic demand from the users on the ground, and 2) calculating the forwarding tables to be used by the UAVs. Two types of UAVs compose the FMN, as illustrated in Fig.~\ref{fig:wise-concept}: Flying Mesh Access Points (FMAPs) and a Gateway (GW) UAV. We model the FMN at time instant $t_k = k \cdot \Delta t, k \in N_0$ and $\Delta t \in \mathbb{R}$ as a directed graph $G(t_k)=(V, E(t_k))$, where $V\in\{1, ..., N\}$ is the set of UAVs and $E(t_k) \subseteq  V \times V $ is the set of directional communications links between any two UAVs $i$ and $j$, at $t_k$, where $i, j \in V$. The channel between any two UAVs is modeled by the Friis path loss model. 
The directional wireless communications link exists if and only if the SNR is higher than a threshold $S$. A path is defined as a set of adjacent links connecting UAV $i$ to the GW UAV; multiple paths may be available for UAV $i$ at $t_k$, but only one of them is used.  

\subsection{Interference Model}
Taking into account the IEEE 802.11 MAC protocol, for a packet transmission to be successful neither the transmitter nor the receiver should be interfered by other nodes. Hence, the transmissions on links $(i, j)_{t_k}$ and $(k, l)_{t_k}$ are both successful at $t_k$ if and only if both $i$ and $j$ are outside the interference range of $k$ and $l$ at $t_k$. This is expressed by the Transmitter-Receiver Conflict Avoidance (TRCA) interference model~\cite{Houaidia2017}.\looseness=-1

\begin{figure}[ht]
	\setlength\abovecaptionskip{-0.1\baselineskip}
	\centering
	\includegraphics[scale=0.18]{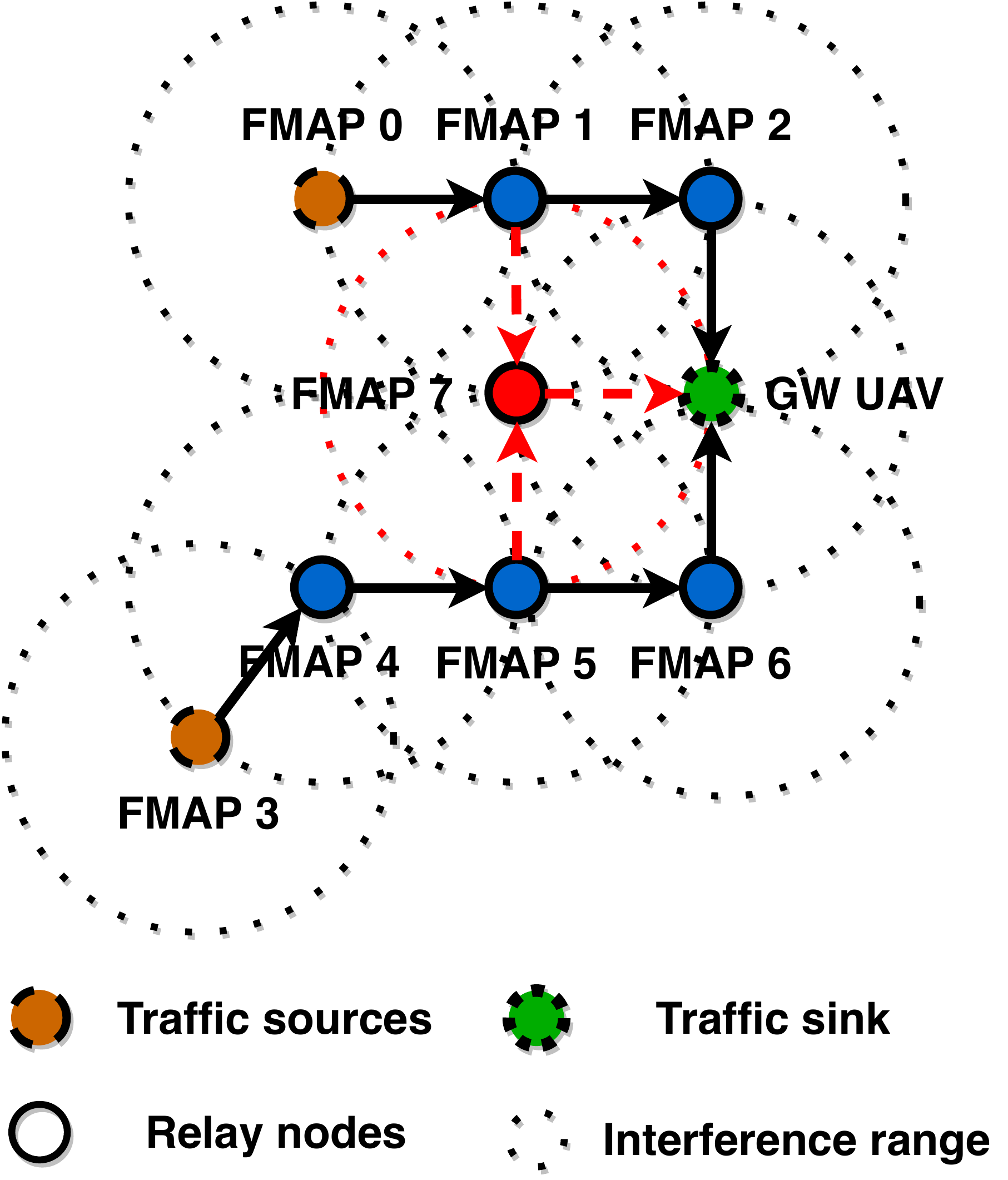}
	\caption{Network graph illustrating the TRCA interference model. If FMAP 7 is used as relay node, the network performance will be reduced up to 50\%, since FMAP 7 is in the interference range of FMAP 1 and FMAP 5.}
	\label{fig:analysis-reference-case}
\end{figure}

In order to demonstrate how the selection of relay nodes affects the network performance, let us analyze a reference case. For the sake of simplicity, we consider the scenario depicted in Fig.~\ref{fig:analysis-reference-case}. It is formed by: 1) two FMAPs generating traffic -- FMAP 0 and FMAP 3; 2) a GW UAV; and 3) six FMAPs able to forward traffic. The interference range of each node is represented by a dashed circumference around that node.  
Firstly, we consider two paths for the flows between the FMAPs generating traffic and the GW UAV: $p_1$$:$$<$$FMAP 0, FMAP 1, FMAP 2, GW UAV$$>$ and $p_2$$:$$<$$FMAP 3, FMAP 4, FMAP 5, FMAP 6, GW UAV$$>$. For these paths, there is no inter-flow interference, hence the throughput achieved by each flow is only limited by the link with the lowest capacity among the ones forming the path. Conversely, if FMAP 7 is chosen to be part of a path $p_1'$$:$$<$$FMAP 0, FMAP 1, FMAP 7, GW UAV$$>$, since FMAP 7 is in the interference range of FMAP 5, the links $<$$FMAP 1, FMAP 7$$>$ and $<$$FMAP 4, FMAP 5$$>$ become mutually interfered. Therefore, the network performance is reduced up to 50\%, when compared with the previous routing configuration.\looseness=-1 

This reference case motivates the definition of an inter-flow interference-aware routing approach to improve the performance of RedeFINE. In fact, by using the Euclidean distance as routing metric, as RedeFINE does, it becomes indifferent selecting FMAP 7, or FMAP 2 and FMAP 6, respectively,  to forward the traffic from FMAP 1 and FMAP 5, since the minimal Euclidean distance is always ensured.\looseness=-1

\section{I2R Routing Metric\label{sec:concept}}
In order to improve the performance of RedeFINE~\cite{Coelho2018}, we propose I2R. It consists of two factors: the distance-aware factor and the inter-flow interference-aware factor. Both factors are fed by the centralized view of the FMN provided by the CS, which knows the future locations of the UAVs and the FMAPs that will be serving the mobile terminals on the ground. Since a strong LoS component characterizes the wireless links between the UAVs flying dozens of meters above the ground, we use Friis path loss to model the links between the UAVs, and estimate the SNR and number of neighboring UAVs. This holistic knowledge avoids the usage of control packets for neighbor discovery and interference estimation. The distance-aware factor is based on the Euclidean distance of the links between each pair of UAVs, at time instant $t_k$, which is denoted by $d_{i,j}(t_k)$. As such, this factor includes the sum of the Euclidean distances of the set of links forming a path $p$, considering in advance the future trajectories that UAVs will follow, which were calculated and pre-defined by the CS to fulfill the traffic demand of the ground users. Using the Euclidean distance as part of the routing metric is compliant with the objective of selecting high-capacity paths, since the link capacity increases as the Euclidean distance decreases, according to the Shannon-Hartley theorem. $d_{i,j} (t_k)$ is normalized to the maximum Euclidean distance among all the usable links of the FMN, at $t_k$. In turn, the inter-flow interference aware factor is a value $\gamma(t_k)$ that is added to the Euclidean distance of the link between UAV $i$ and UAV $j$, at $t_k$. $\gamma_j(t_k)$ represents the number of neighboring nodes of UAV $j$, excluding UAV $i$, at $t_k$. We assume as neighboring nodes the UAVs in carrier-sense range that are generating or forwarding traffic. $\gamma_j(t_k)$ considers that the level of interference is equal either the neighboring nodes are close or far away, as the TRCA model denotes. $\gamma_j(t_k)$ is normalized to the maximum number of neighbors that any UAV composing the FMN has at $t_k$.\looseness=-1

The path cost using I2R is defined in Eq.~\eqref{eq:routing-metric}, where $0 \leq \alpha \leq 1$ is a tunable parameter that weights the influence of the distance-aware and interference-aware factors. To calculate the path between any UAV and the GW UAV, the Dijkstra's algorithm~\cite{skiena1990} is used. Considering the reference case depicted in Fig.~\ref{fig:analysis-reference-case}, I2R uses the factor $\gamma$ to increase the cost of the links $<FMAP 1, FMAP 7>$ and $<FMAP 5, FMAP 7>$, since FMAP 7 is in the interference range of FMAP 1 and FMAP 5.\looseness=-1 

\begin{equation}
	I2R = (1-\alpha)\times\sum_{\forall (i,j) \in p}d_{i,j}(t_k) + \alpha \times \sum_{\forall j \in p} \gamma_{j}(t_k)
	\label{eq:routing-metric}
\end{equation}

\section{\label{sec:performance-evaluation}Performance Evaluation}
The performance evaluation of the I2R routing metric is presented in this section, including the simulation setup, the simulated scenarios, and the performance metrics considered.

\subsection{Simulation Setup}
The ns-3 simulator was used to evaluate the proposed routing metric in complex networking scenarios formed by an FMN composed of 1 GW UAV and 20 FMAPs. In each UAV, a Network Interface Card (NIC) was configured in Ad Hoc mode, using the IEEE 802.11ac standard in channel 50, which allows \SI{160}{\mega\hertz} channel bandwidth. The data rate was defined by the \textit{IdealWifiManager} mechanism. The wireless links between the UAVs were modeled by Friis path loss; only links with SNR above \SI{5}{\deci\bel} were considered as usable. The transmission power of the NICs was set to \SI{0}{\deci\belmilliwatt}. 

One IEEE 802.11ac spatial stream was used for the wireless links. With one spatial stream, the data corresponding to the maximum Modulation and Coding Scheme (MCS) index is \SI{780}{Mbit/s}, considering \SI{800}{\nano\second} Guard Interval. Taking into account the dimensions of the simulated scenarios, we assume an average number of 2 hops between the FMAPs generating traffic and the GW UAV; this results in $\frac{\frac{780}{N_{tx}}}{2}\SI{}{Mbit/s}$ for the maximum achievable data rate per flow, where $N_{tx}$ denotes the number of FMAPs generating traffic. Based on that, the maximum offered load for each scenario was set to 75\% of the maximum achievable data rate per flow, for a total number of FMAPs generating traffic between 5 and 10. The traffic generated was UDP with arrival process modeled as Poisson, for a constant packet size of 1400 bytes; the traffic generation was only triggered after \SI{30}{\second} of simulation, in order to ensure a stable state. In addition, different values for the tunable parameter $\alpha$, between 0.2 and 1, were considered. The Controlled Delay (CoDel) algorithm~\cite{nichols2012}, which is a Linux-like queuing discipline that considers the time that packets are held in the transmission queue to decide when to discard packets, was used; it allows to mitigate the bufferbloat problem. The default parameters of CoDel in ns-3 were employed.

\subsection{\label{sec:scenarios}Simulation Scenarios}
Five scenarios, in which the UAVs were moving according to the Random Waypoint Mobility (RWM) model, were generated to evaluate the performance of RedeFINE using the I2R routing metric in typical crowded events. Under the RWM model, each UAV chooses a random destination and a speed uniformly distributed between a minimum and a maximum value. Then, the UAV moves to the chosen destination at the selected speed; upon arrival, the UAV stops for a specified period of time and repeats the process for a new destination and speed~\cite{Camp2002}.
Since I2R relies on knowing in advance the movements of the UAVs, instead of generating the random movements during the ns-3 simulation, we used BonnMotion~\cite{aschenbruck2010bonnmotion}, which is a mobility scenario generation tool. BonnMotion was set to create Random Waypoint 3D movements for 21 nodes (20 FMAPs and 1 GW UAV) within a box of dimensions \SI{80}{\meter} $\times$ \SI{80}{\meter} $\times$ \SI{25}{\meter}  during \SI{160}{\second}, considering a velocity between \SI{0.5}{m/s} and \SI{3}{m/s} for the UAVs. These scenarios were used to calculate in advance the forwarding tables and the instants they shall be updated in the UAVs. Both the forwarding tables and the generated scenarios were finally imported to ns-3 with a sampling period of \SI{1}{\second}. To employ mobility to the UAVs, based on the generated scenarios, the \textit{WaypointMobilityModel} model of ns-3 was used.

\subsection{Performance Metrics}
RedeFINE using the I2R routing metric was evaluated against two state of the art distributed routing protocols representative of the reactive and proactive routing paradigms -- Ad Hoc On Demand Distance Vector (AODV)~\cite{aodv-etx-repository} and Optimized Link State Routing Protocol (OLSR)~\cite{olsr-etx-repository}, respectively, using the ETX routing metric, -- and against RedeFINE using Airtime and the Euclidean distance routing metric. ETX is a link quality-based routing metric that represents the expected number of transmissions required to send a packet over a link, including retransmissions. Airtime, which is the default routing metric specified in the IEEE 802.11s standard~\cite{Hiertz2010}, expresses the amount of channel resources consumed for transmitting a frame over a link. Since the theoretical calculation of the Airtime resulting costs is not straightforward, we exported them from ns-3, by running previous simulations for each one of the generated scenarios. Afterwards, we employed the Dijkstra's algorithm to find the shortest paths between each UAV and the GW UAV, considering a sampling period of \SI{1}{\second} and the corresponding routing metric. The Airtime routing metric was used in our evaluation to ensure that I2R is able to overcome a metric that uses real measurements to estimate data rate, overhead, and frame error rate of the communications links. 

Our performance evaluation considers two performance metrics:
\begin{itemize}
	\item \textbf{Aggregate throughput} -- The mean number of bits received per second by the GW UAV.
	\item \textbf{End-to-end delay} -- The mean time taken by the packets to reach the application layer of the GW UAV since the instant they were generated at a given FMAP, measured at each second, including queuing, transmission, and propagation delays. 
\end{itemize}

\subsection{Simulation Results\label{sec:simulation-results}}
The simulation results are presented in this section. The results were obtained after 20 simulation runs, using different seeds, for each experimental combination, including different $\alpha$ values and different number of FMAPs generating traffic. The results are expressed using mean values, considering five random scenarios, as stated in Subsection~\ref{sec:scenarios}. They are represented by means of the Cumulative Distribution Function (CDF) for the end-to-end delay and by the complementary CDF (CCDF) for the aggregate throughput, including the values for the $25^{th}$, $50^{th}$, and $75^{th}$ percentiles. The CDF $F(x)$ represents the percentage of simulation time for which the mean end-to-end delay was lower or equal to $x$, while the CCDF $F'(x)$ represents the percentage of simulation for which the mean aggregate throughput was higher than $x$. 
Finally, the influence of the tunable parameter $\alpha$ on the FMN performance is also evaluated. 

\begin{figure}[ht]
		
	\centering
	\subfloat[End-to-end delay Cumulative Distribution Function (CDF).]{
		\includegraphics[scale=0.13, keepaspectratio]{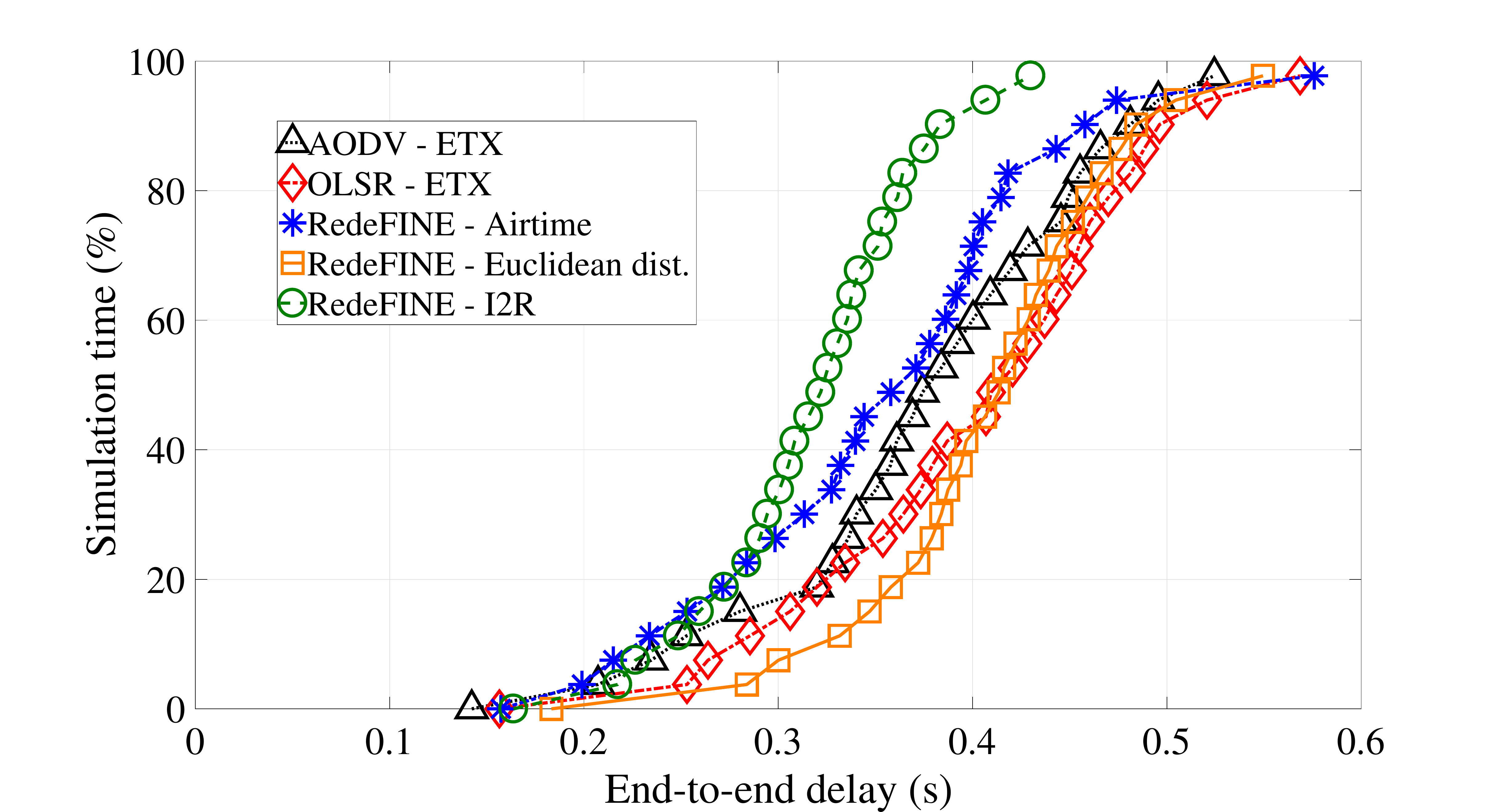}
		\label{fig:delay-cdf-5txnodes}}
	\hfill
	\subfloat[Throughput Complementary Cumulative Distribution Function (CCDF).]{
		\includegraphics[scale=0.13, keepaspectratio]{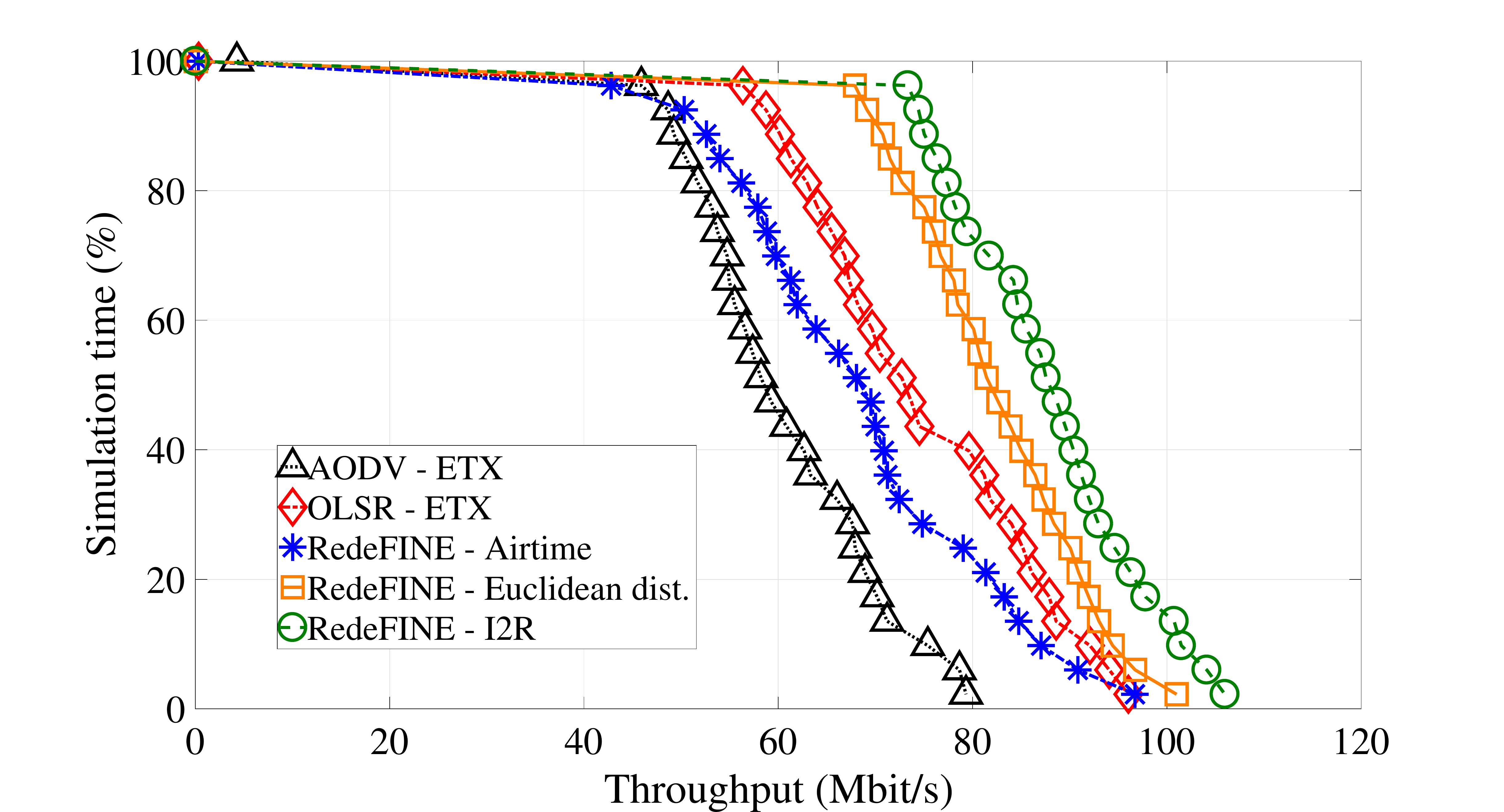}
		\label{fig:throughput-ccdf-5txnodes}}
	\hfill
	\subfloat[The $25^{th}$, $50^{th}$, and $75^{th}$ percentiles of both the throughput CCDF and end-to-end delay CDF.]{
		\includegraphics[scale=0.13, keepaspectratio]{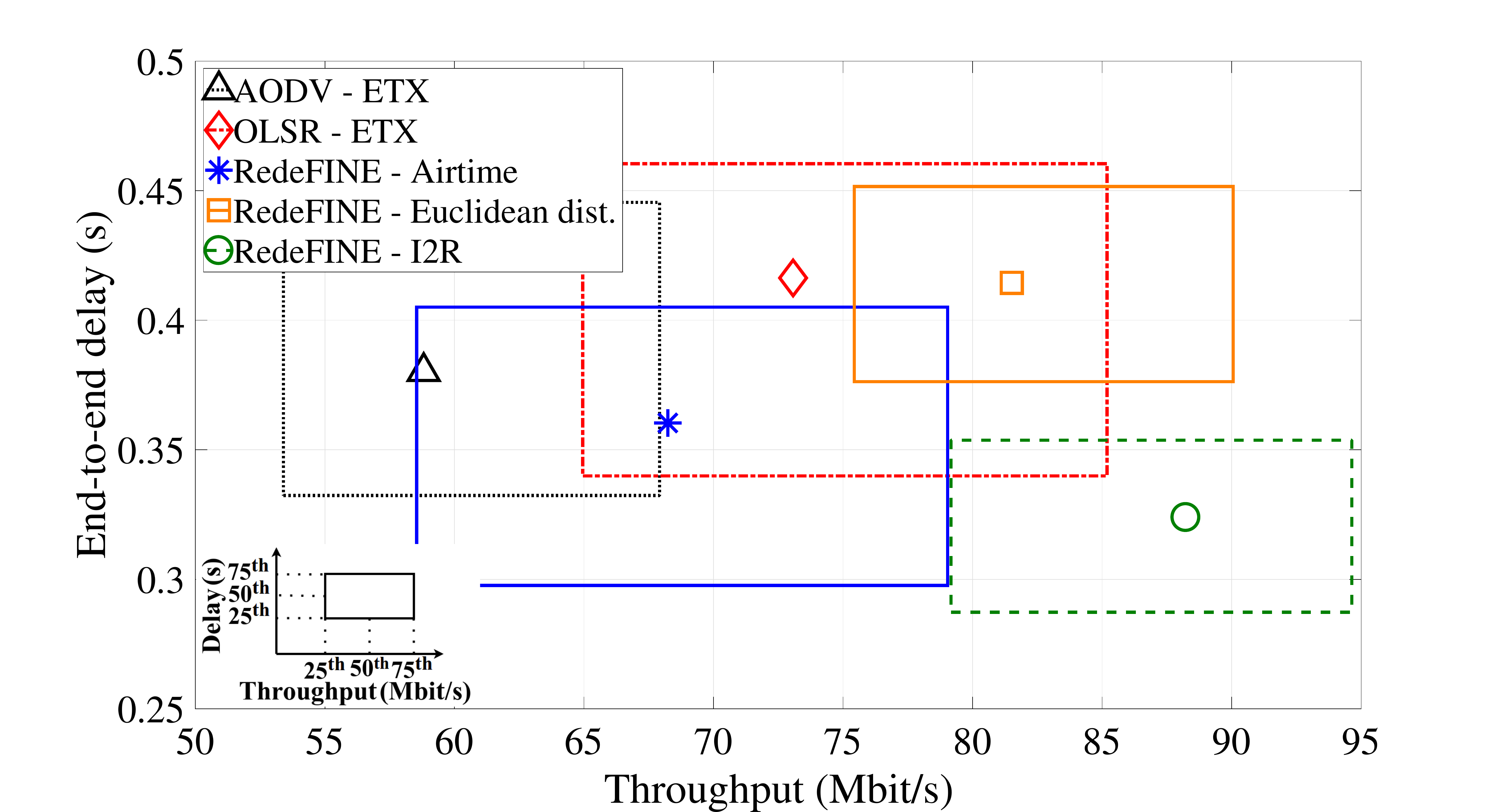}
		\label{fig:throughput-delay-5txnodes}}
	\caption{Results for throughput and end-to-end delay in the GW UAV. The results were obtained considering 5 FMAPs generating traffic, and $\alpha=1$.}
	\label{fig:results_5txnodes}
\end{figure}

\begin{figure}[ht]
	\centering
	\subfloat[End-to-end delay Cumulative Distribution Function (CDF).]{
		\includegraphics[scale=0.13, keepaspectratio]{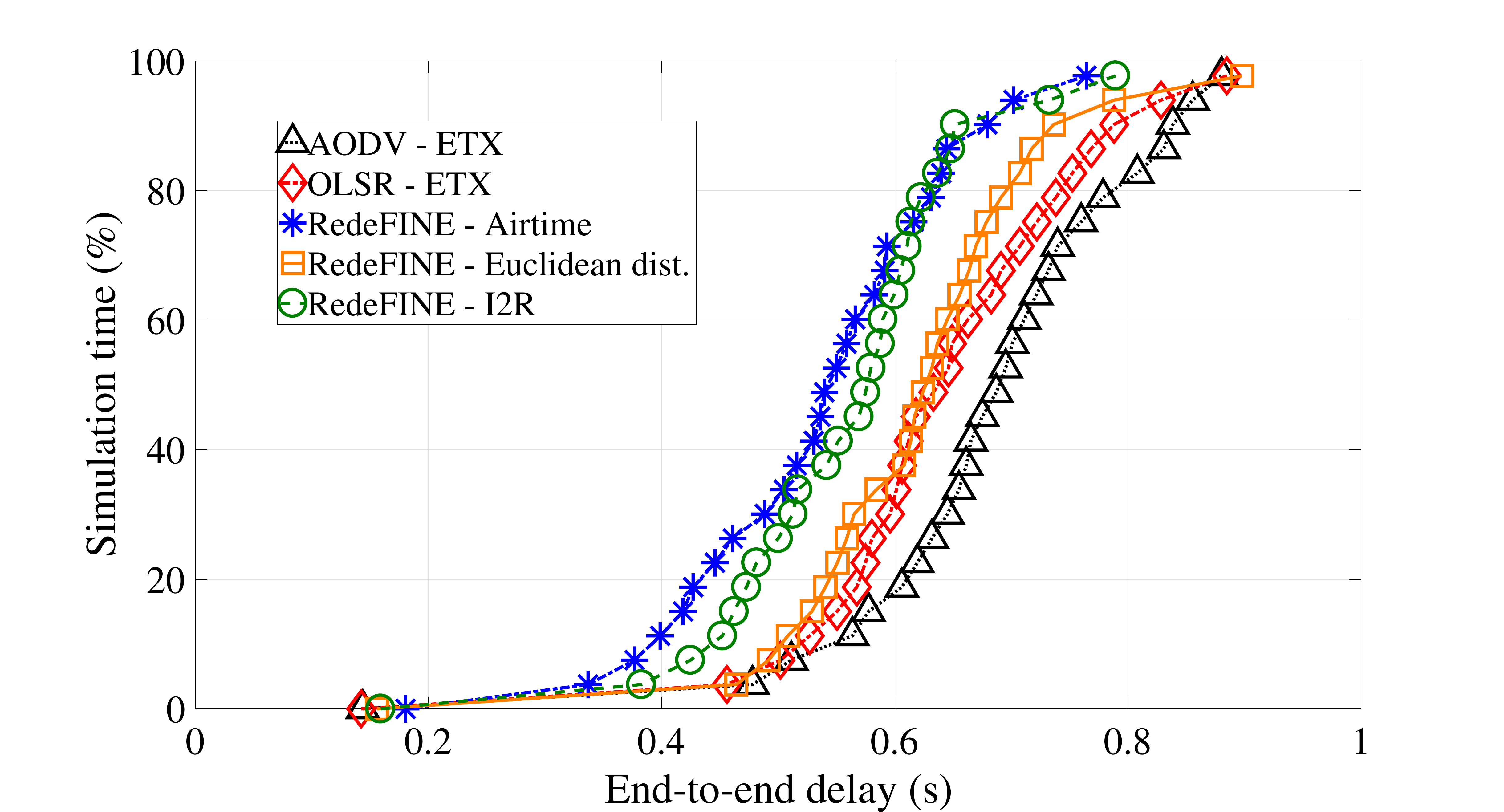}
		\label{fig:delay-cdf-10txnodes}}
	\hfill
	\subfloat[Throughput Complementary Cumulative Distribution Function (CCDF).]{
		\includegraphics[scale=0.13, keepaspectratio]{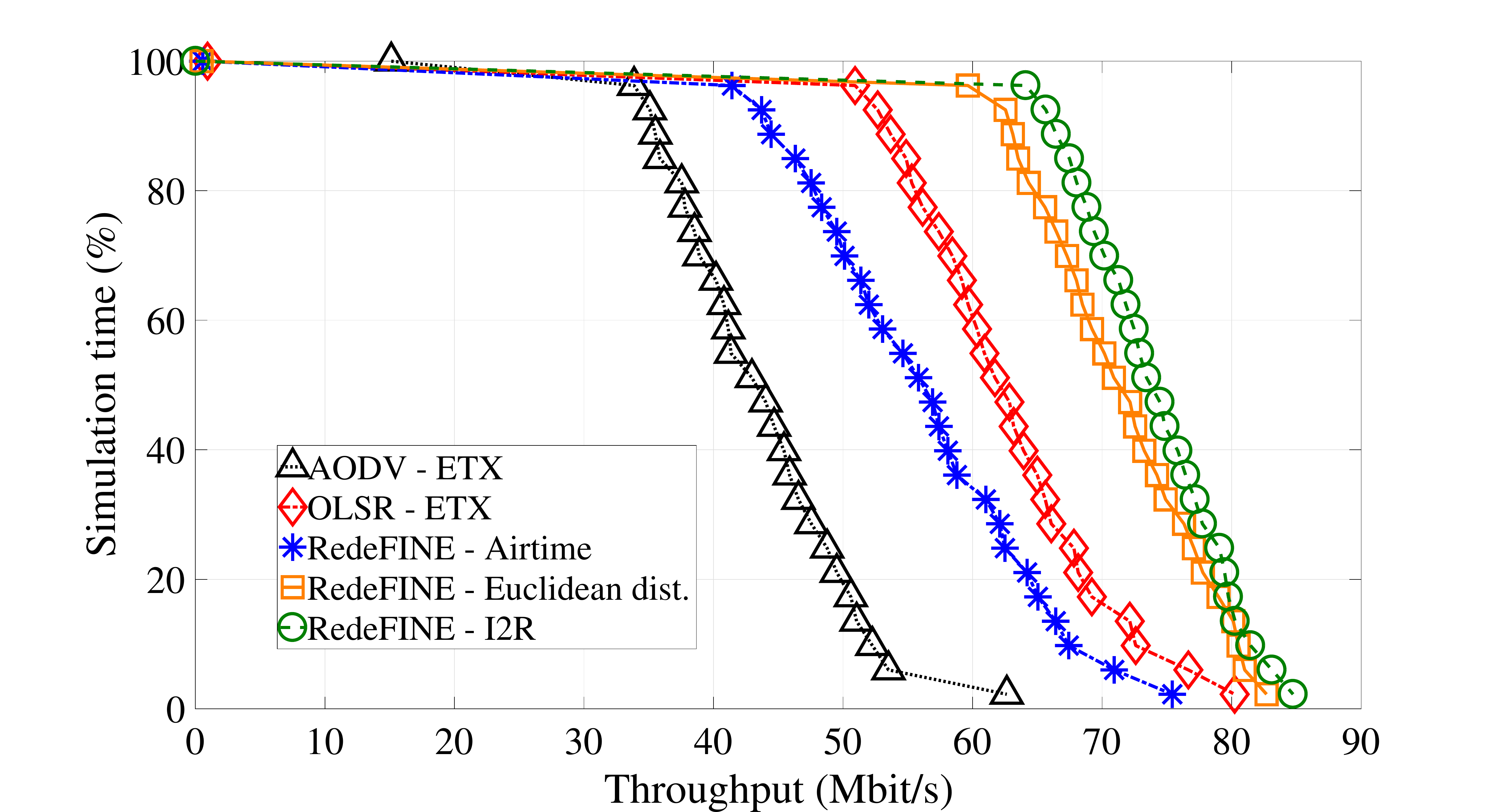}
		\label{fig:throughput-ccdf-10txnodes}}
	\hfill
	\subfloat[The $25^{th}$ , $50^{th}$, and $75^{th}$ percentiles of both the throughput CCDF and end-to-end delay CDF.]{
		\includegraphics[scale=0.13, keepaspectratio]{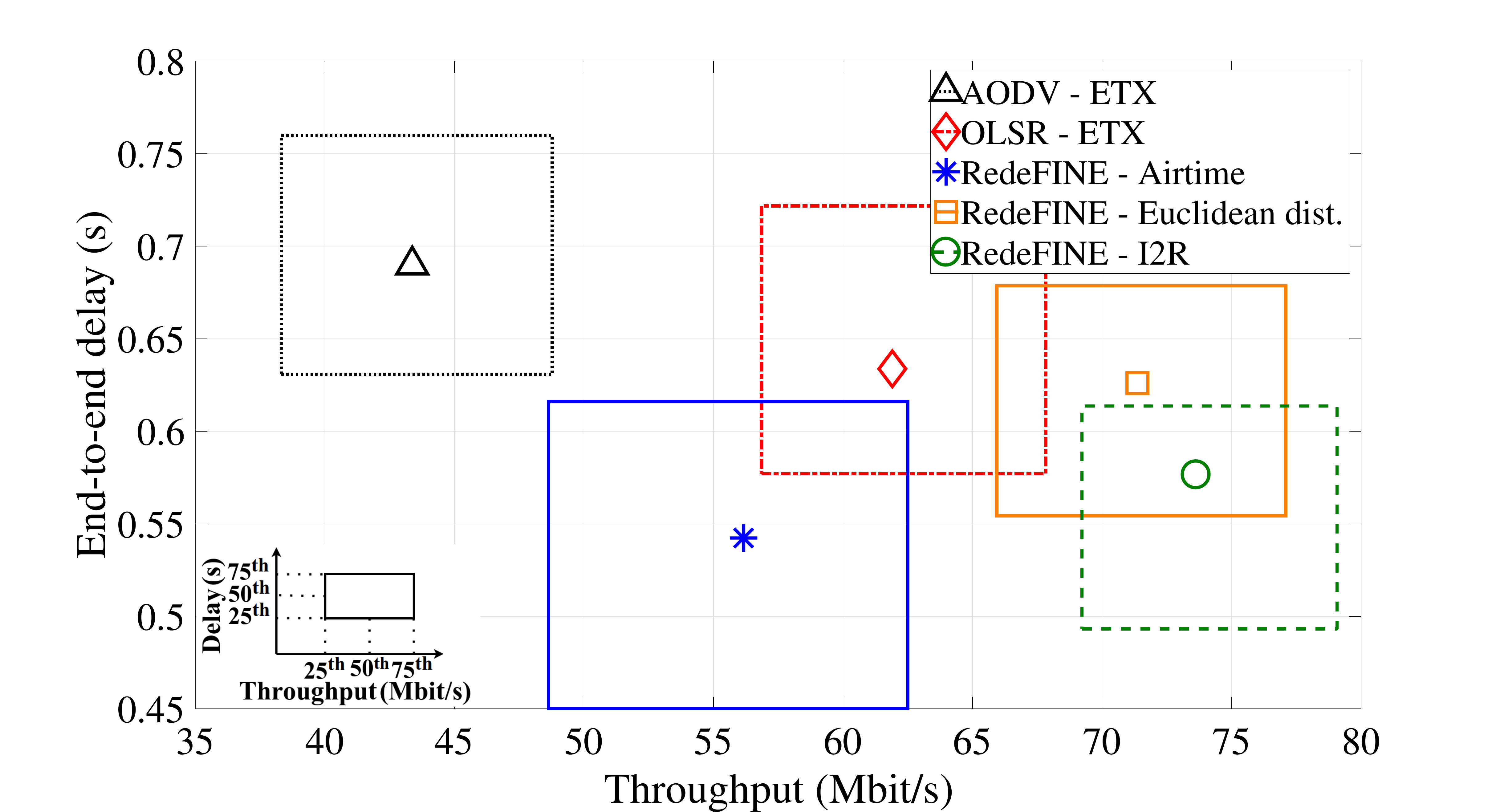}
		\label{fig:throughput-delay-10txnodes}}
	\caption{Results for throughput and end-to-end delay in the GW UAV. The results were obtained considering 10 FMAPs generating traffic, and $\alpha=1$.}
	\label{fig:results_10txnodes}
\end{figure}

When 5 FMAPs are used as traffic sources (cf. Fig.~\ref{fig:results_5txnodes}), the usage of the I2R routing metric improves the end-to-end delay achieved by RedeFINE using the Euclidean distance in approximately 22\%, OLSR and AODV using ETX in 21\% and 15\%, respectively, and RedeFINE using Airtime in 10\%. These values are obtained considering the mean end-to-end delay of the packets received in the GW UAV for the different solutions. The outperforming results of RedeFINE using the I2R routing metric are justified by the selection of paths formed by UAVs with reduced number of neighbors that are generating or forwarding traffic. Regarding the total amount of bits received in the GW UAV, RedeFINE using the I2R routing metric provides a gain up to 7\% when compared with RedeFINE using the Euclidean distance. In turn, when compared with AODV and OLSR using the ETX routing metric, and RedeFINE using Airtime, the gains are even more relevant: approximately 45\%, 17\%, and 28\%, respectively.  
When 10 FMAPs are generating traffic (cf. Fig.~\ref{fig:results_10txnodes}), RedeFINE using the I2R routing metric improves end-to-end delay in approximately 10\% with respect to RedeFINE using the Euclidean distance, while the gain over AODV and OLSR using ETX is approximately 18\% and 13\%, respectively. The gain in end-to-end delay of I2R over the Airtime routing metric applied to RedeFINE is negligible.
Regarding the total amount of bits received in the GW UAV, the gain of RedeFINE using I2R over OLSR using ETX is still approximately 18\%, and over RedeFINE using the Euclidean distance is negligible ($\approx$ 4\%). Conversely, the gain over AODV using ETX is increased to approximately 68\%, while with respect to RedeFINE using Airtime it is approximately 31\%. 

The relation between aggregate throughput and end-to-end delay for the different combinations of protocols and routing metrics is depicted in Fig.~\ref{fig:throughput-delay-5txnodes} and Fig.~\ref{fig:throughput-delay-10txnodes}, where the $25^{th}$, $50^{th}$, and $75^{th}$ percentiles of both the throughput CCDF and delay CDF are represented. Overall, the gains in end-to-end delay and throughput of RedeFINE using I2R are reduced when the number of transmission FMAPs increases. 
The performance evaluation carried out allowed to conclude that I2R selects preferably as relay nodes the FMAPs that are also sources of traffic; for instance, in a scenario where 5 FMAPs are generating traffic, if any of these FMAPs need a relay to reach the GW UAV, then I2R will give preference to any of the remaining 4 FMAPs that are generating traffic, thus avoiding that a sixth FMAP introduces interference in the FMN. This effect is faded when the number of FMAPs in the FMN increases.    

\begin{figure}[ht]
	\setlength\abovecaptionskip{-0.1\baselineskip}
	\centering
	\includegraphics[scale=0.16]{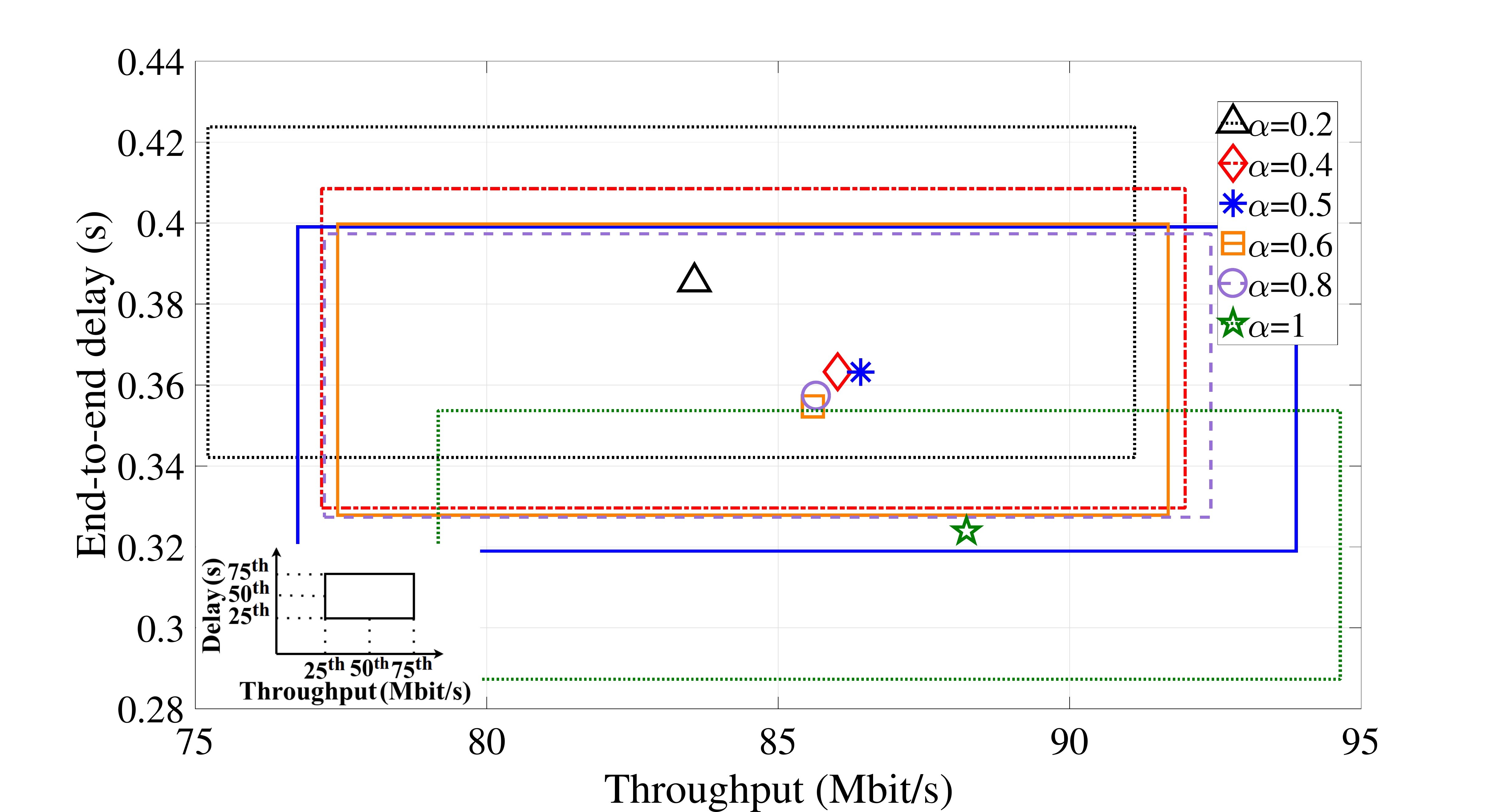}
	\caption{The $25^{th}$, $50^{th}$, and $75^{th}$ percentiles of both the throughput CCDF and delay CDF, considering different $\alpha$ values. The results consider 5 FMAPs generating traffic.}
	\label{fig:thorughput-delay-different-alpha}
\end{figure}

Regarding the tunable parameter $\alpha$, it must be set to a value close to 1 for higher throughput and lower end-to-end delay values. As $\alpha$ decreases, the performance worsens, as exacerbated by $\alpha=0.2$, in Fig.~\ref{fig:thorughput-delay-different-alpha}. This demonstrates how the selection of paths formed by the minimum number of neighboring UAVs in carrier-sense range contributes to improve the performance of an FMN, rather than the selection based only on the Euclidean distance.

The simulation results show that I2R improves the performance of an FMN, especially regarding end-to-end delay. It outperforms the link quality-based routing metrics ETX and Airtime, and the Euclidean distance routing metric considered in the previous version of RedeFINE. This can be justified by the fact that the ETX and Airtime routing metrics provide estimations based on small probe packets. As consequence, the link loss rate is underestimated, especially if high data rates are used, as IEEE 802.11ac does. When the Euclidean distance is used as routing metric, it may result in multiple nodes close to each other transmitting simultaneously, causing inter-flow interference within the FMN. In the worst case scenario, I2R enables better end-to-end delay than the Euclidean distance routing metric for the same total amount of bits received in the GW UAV. This is our main contribution. 

RedeFINE using the I2R routing metric represents an inter-flow interference-aware approach that estimates in advance the number of neighboring UAVs without using control packets for neighbor discovery, link sensing, and interference estimation, based on the decisions performed by the CS that defines the FMN topology. Since forwarding tables are computed centrally, the computational power on board of the UAVs can be reduced. 

In its current version, I2R does not consider the amount of traffic generated by the UAVs and intra-flow interference. In addition, the improvement of I2R for an FMN composed of multiple GW UAVs is worthy to be considered. These are aspects left for future work.

\section{Conclusions\label{sec:conclusions}}
We presented I2R, an interference-aware  routing  metric  specially  designed for centralized  routing in FMNs. I2R  does  not  require  any  control  packets and  enables  the  configuration  of  paths  with  minimal Euclidean distance  formed  by  UAVs  with  the  lowest  number  of  neighbors in carrier-sense range, thus minimizing inter-flow interference in the FMN. Based on simulation results for five random scenarios and different number of FMAPs generating traffic, we demonstrated the superior performance of I2R with respect to the routing metrics ETX, Airtime, and Euclidean distance. Simulation results show that I2R improves the performance of RedeFINE using the Euclidean distance routing metric up to 22\% in end-to-end delay and up to 7\% in throughput. As future work, we will improve I2R to take into account the amount of traffic generated by the UAVs. In addition, intra-flow interference and selection between multiple GW UAVs are worthy to be considered.             

\section*{Acknowledgments}
This work is financed by National Funds through the Portuguese funding agency, FCT -- Fundação para a Ciência e a Tecnologia within project: UID/EEA/50014/2019. This work is also part of the FCT WISE project POCI-01-0145-FEDER-016744, financed by the ERDF -- European Regional Development Fund through the Operational Programme for Competitiveness and Internationalisation -- COMPETE 2020 Programme, and by National Funds through the Portuguese funding agency, FCT. The  first  author also thanks  the  funding from FCT under the PhD grant SFRH/BD/137255/2018.

\bibliographystyle{IEEEtran}
\bibliography{IEEEabrv,references}

\end{document}